\documentclass[%
reprint,
amsmath,amssymb,
aps,
]{revtex4-1}

\usepackage{graphicx}
\usepackage{dcolumn}
\usepackage{bm}
\usepackage{hyperref}
\usepackage{siunitx}
\usepackage{amsmath}
\usepackage{mathtools}

\newcommand{\beginsupplement}{%
	\setcounter{table}{0}
	\renewcommand{\thetable}{S\arabic{table}}%
	\setcounter{figure}{0}
	\renewcommand{\thefigure}{S\arabic{figure}}%
	\setcounter{subsection}{0}
	\renewcommand{\thefigure}{S\arabic{subsection}}%
}

\begin{document}
	
	
	\title{A nanoelectromechanical position-sensitive detector with picometer resolution}
	
	\author{Miao-Hsuan Chien}
	\affiliation{Institute of Sensor and Actuator Systems, TU Wien, 1040 Vienna, Austria.}
	\author{Johannes Steurer}
	\affiliation{Institute of Sensor and Actuator Systems, TU Wien, 1040 Vienna, Austria.}
	\author{Pedram Sadeghi}
	\affiliation{Institute of Sensor and Actuator Systems, TU Wien, 1040 Vienna, Austria.}
	\author{Nicolas Cazier}
	\affiliation{Institute of Sensor and Actuator Systems, TU Wien, 1040 Vienna, Austria.}
	\author{Silvan Schmid}
	\email{silvan.schmid@tuwien.ac.at}
	\affiliation{Institute of Sensor and Actuator Systems, TU Wien, 1040 Vienna, Austria.}
	
	\date{\today}
	
	\begin{abstract}
		
		Sub-nanometer displacement detection lays the solid foundation for critical applications in modern metrology. In-plane displacement sensing, however, is mainly dominated by the detection of differential photocurrent signals from photodiodes, with resolution in the nanometer range. Here, we present an integrated in-plane displacement sensor based on a nanoelectromechanical trampoline resonator. With a position resolution of \SI{4}{\pico\meter\per\sqrt{\hertz}} for a low laser power of \SI{85}{\micro\watt} and a repeatability of \SI{2}{\nano\meter} after 5 cycles of operation as well as good long-term stability, this new detection principle provides a reliable alternative for overcoming the current position detection limit in a wide variety of research and application fields.
		
	\end{abstract}
	
	\keywords{In-plane displacement sensor, Position-sensitive detector, Nanoelectromechanical resonators, Photothermal effect}
	
	\maketitle
	
	\section{Introduction}
	\begin{figure*}
		\centering
		\includegraphics[width=\textwidth]{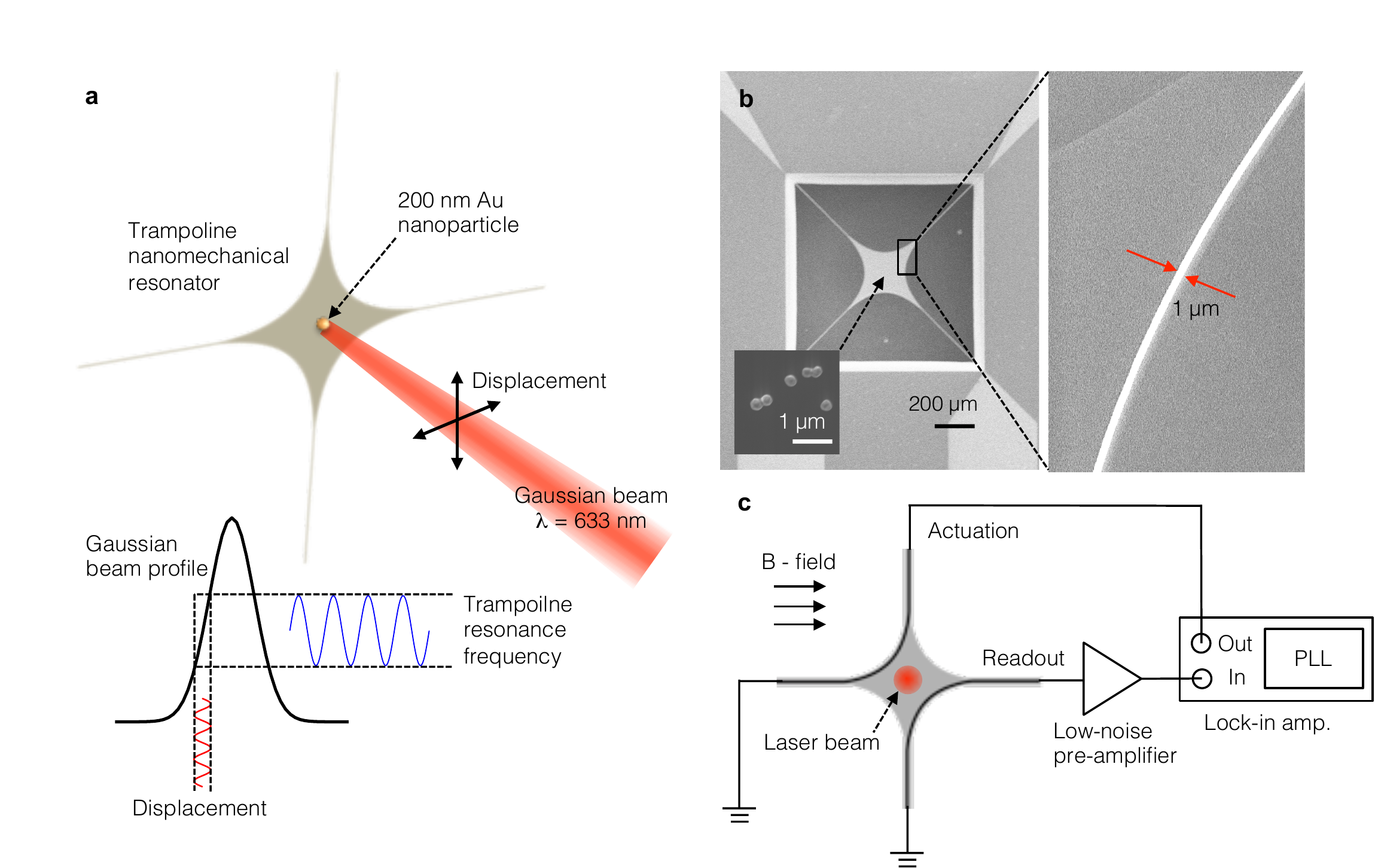} 
		\caption{(a) Working principle of the nanoelectromechanical position-sensitive detector (NEMS-PSD). Displacement of the light beam could be detected by the shift of the trampoline resonance frequency as a result of the change in absorption of gold nanoparticle. (b) SEM image of trampoline nanomechanical resonator with readout and actuation electrodes, with a close-up SEM image of the gold electrode in the marked area. Both electrodes are \SI{1}{\micro\meter} wide and following the profile of the trampoline with approximately \SI{1}{\micro\meter} spacing to the edge. The electrodes have a resistance of \SI{150}{\ohm} resulting in a Johnson noise of below \SI{2}{\nano\volt\per\sqrt{\hertz}}. 200 nm Au nanoparticles are distributed on the center area of the trampoline, as shown in the SEM image in the inset. (c) Detection scheme of the NEMS-PSD. The trampoline features two gold electrodes, one for inductive readout and the other for Lorentz force actuation. To obtain an optimal signal-to-noise ratio an enhanced Halbach array is used to create a static magnetic field of approximately \SI{1}{\tesla}. The magnetomotive readout current is amplified by a homemade low-noise pre-amplifier. The resonance frequency is then tracked with a phase-locked loop (PLL).
		}
		\label{fig:fig1} 
	\end{figure*}

	High-performance position sensing is a substantial corner stone for challenging applications such as state-of-the-art atomic force microscopy for molecular- and mechano-biology \cite{dufrene2017imaging,krieg2019atomic}, nanomechanical transduction and sensing \cite{knobel2003nanometre,naik2009towards,hanay2012single,sage2015neutral}, and experimental physics such as the tracking of single electron spins and trapped ions \cite{biercuk2010,grinolds2013nanoscale}. Position-sensitive detectors (PSD) that can measure the position of a light spot are an integral part of modern metrology. The two most common PSD designs are based on segmented or lateral effect sensors. The former PSDs consist of multiple sensor segments each giving its own photocurrent, while the latter is based on a single photodetector element.
	Typical segmented-quadrant position-sensitive detectors have a good position resolution of the order of \SIrange{10}{100}{\nano\meter\per\sqrt{\hertz}} for light powers of \SIrange{10}{100}{\micro\watt} \cite{makynen2000position}. Due to their quick response time and large operation bandwidth, they are dominating the commercial atomic force microscope market. Each detector segment is separated by a gap. The intensity profile of a spot is in general very nonlinear, which directly results in a strong nonlinear position response if the spot is not perfectly centered. 
	In contrast, lateral effect position-sensitive detectors have no gaps and give positional information independent of beam shape, size, and intensity profile. Lateral effect PSDs have a good lateral resolution in the range of a few \SI{}{\nano\meter\per\sqrt{\hertz}}, however typically have a slower response speed than quadrant detectors.  \cite{andersson2008position,makynen2000position}. 
	For both PSDs, as a characteristic of semiconductor photodetectors, the dark-current noise can limit the detector sensitivity to several orders higher than the shot-noise-limit without any considerations of external artifacts, setting an upper limit for the sensitivity of photodiode-based PSDs \cite{azaryan2019position,makynen2000position}.
	
	Besides the common photodiode-based PSDs, it has been shown that a spatial resolution of \SI{25}{\pico\meter\per\sqrt{\hertz}} can be achieved by direct transmission of a Gaussian beam through a slit \cite{haddad2008gaussian}. This transmission based experiment however requires a photodetector behind the slit, which renders it impractical for the use in more general applications. Recently, a fundamentally new approach based on the directional scattering of a laser spot on a silicon nanoantenna has been demonstrated \cite{bag2019}. This nanophotonic displacement sensor has reached a position resolution in the nanometer range.
	
	Here, we present a nanoelectromechanical position-sensitive detector (NEMS-PSD) with a position resolution in the picometer range. Recently, similar nanoelectromechanical resonators have demonstrated unprecedented sensitivity for radiation \cite{piller2019} as well as single nanoparticle and molecule absorption detection \cite{schmid2014low,larsen2013photothermal,chien2018}. The NEMS-PSD principle is based on the highly beam position dependent photothermal heating of plasmonic Au nanoparticles that are placed on top of a silicon nitride trampoline resonators, as schematically depicted in Figure~\ref{fig:fig1}a. A scanning electron microscope image of the trampoline resonator is shown in Figure~\ref{fig:fig1}b. The NEMS-PSD is transduced electrodynamically \cite{cleland1996,cleland1999,venstra2009}, as schematically depicted in Figure~\ref{fig:fig1}c. More details of the NEMS-PSD design can be found in the supplementary information. During operation, due to the Gaussian power distribution of the beam, the displacement of the laser beam changes the power absorbed by the nanoparticles, resulting in a detectable frequency shift of the temperature sensitive nanoelectromechanical resonator. In this way, the presented NEMS-PSD reached a position resolution of \SI{4}{\pico\meter\per\sqrt{\hertz}} for a laser power of \SI{85}{\micro\watt}.
	
	\begin{figure*}
		\centering
		\includegraphics[width=\textwidth]{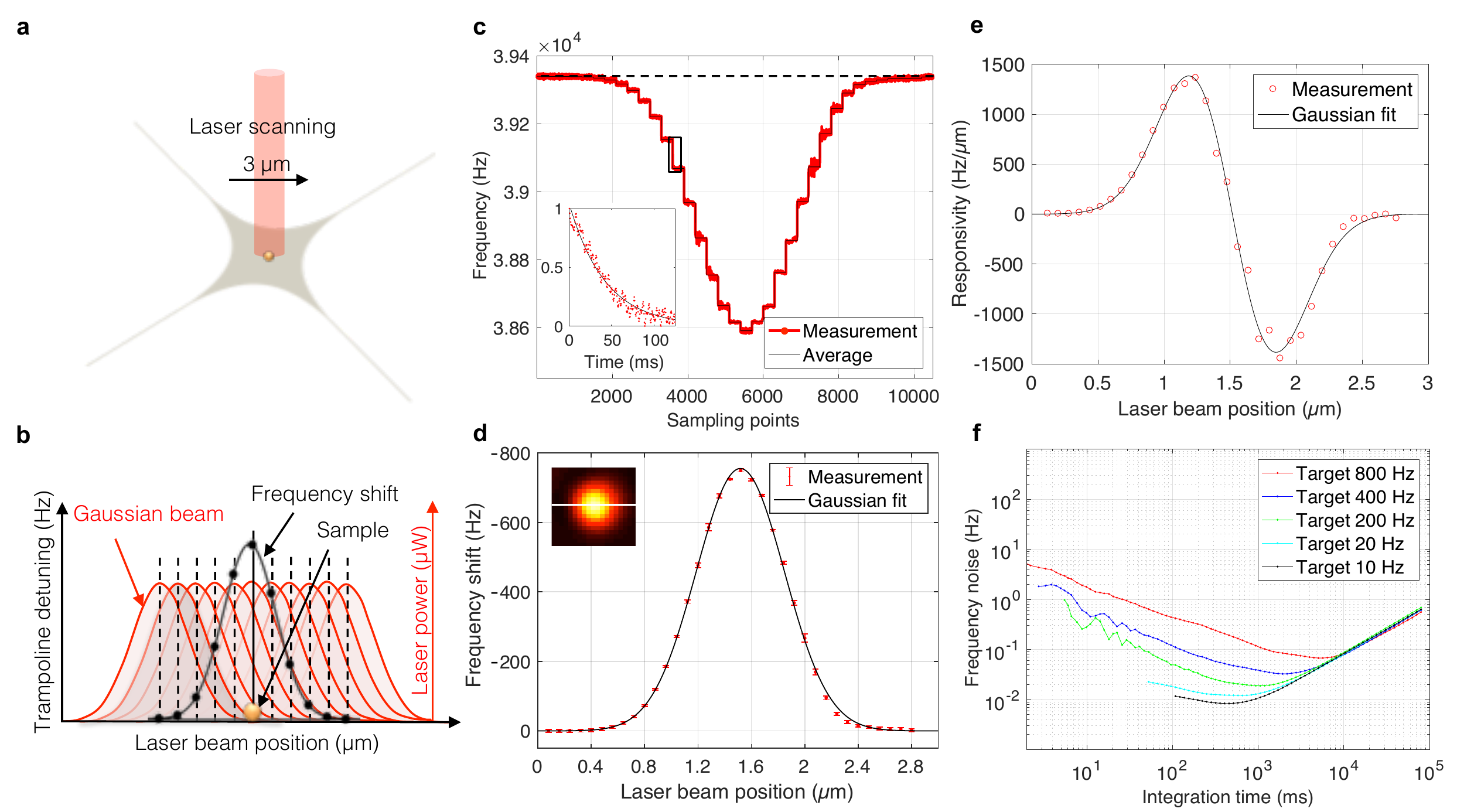} 
		\caption{(a) Schematic of the long-range scanning of the single nanoparticle absorber to investigate the position-dependent displacement responsivity. (b) Schematic of the formation of Gaussian frequency shift profile from the Gaussian beam. As the beam scans through the nanoparticle, the Gaussian power distribution results in a Gaussian absorption profile and thus a Gaussian frequency shift. (c) Resonance frequency tracked by the PLL with a target bandwidth of \SI{800}{\hertz}. The reference frequency is indicated by the dashed line. The mean frequency of each scan step is marked by the solid line. The rise time is extracted from the fit indicated in the inset. (d) Extracted frequency shift of each scan point averaged over the dwell time. The error bars represent the respective standard deviation. The inset shows the full 2D scan (e) The responsivity for the different beam positions assuming piece-wise linearity, with the first derivative of the Gaussian fit from (d) indicated with a solid line. (f) Allan deviation for different target bandwidths.}
		\label{fig:fig3} 
	\end{figure*}
	
	\begin{figure}
		\centering
		\includegraphics[width=0.5\textwidth]{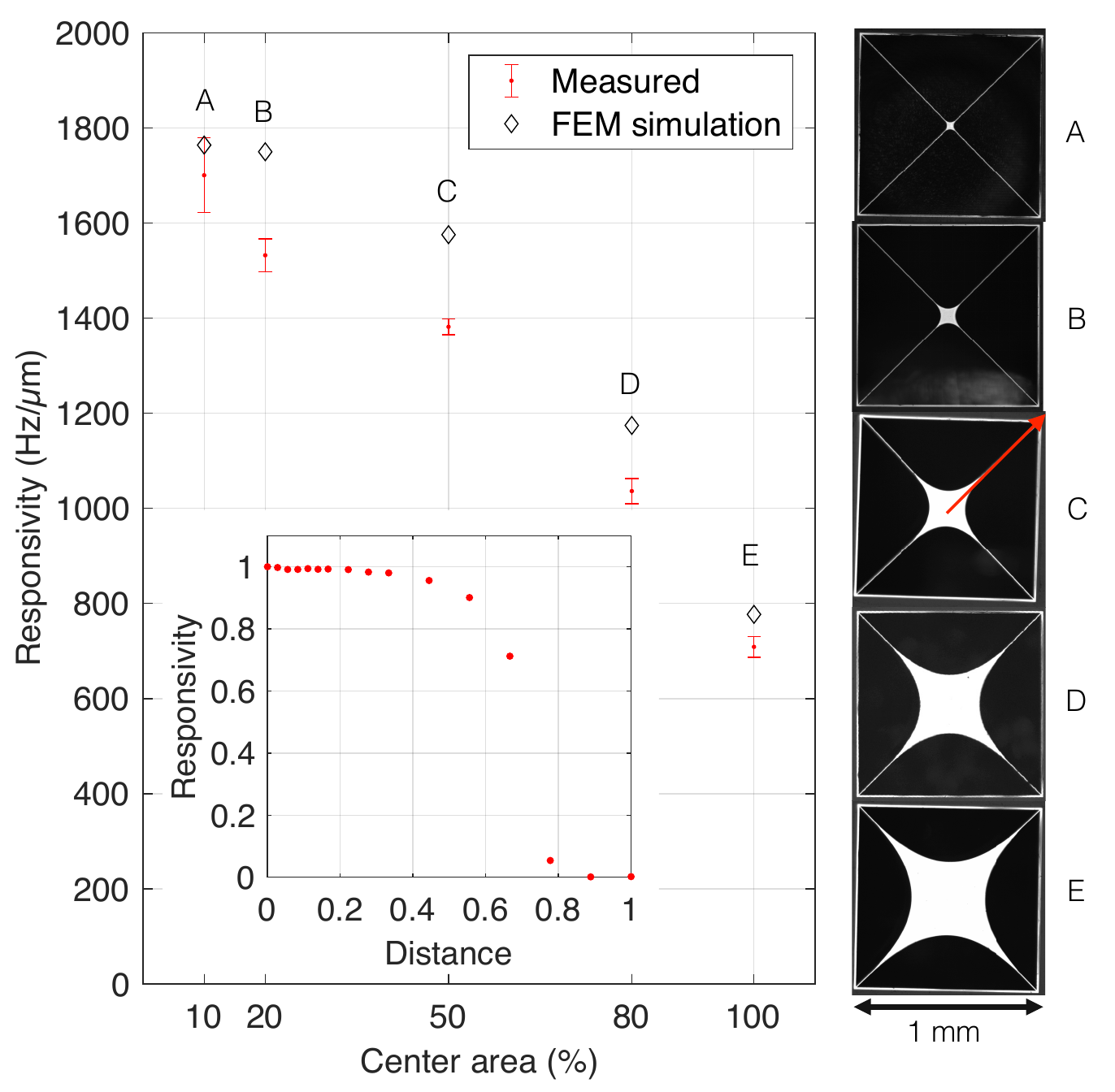} 
		\caption{The measured and simulated responsivity for trampoline resonators with different geometries, indicated as type A to type E with corresponding microscopic images. The center area is marked in percentage, which is a factor of the scaling, with type E representing maximum scaling of 100\%. The FEM-simulated distance dependency of the responsivity is plotted in the inset. All values are normalized for better visualization. The direction of the distance is indicated with the red arrow in the image of the type C trampoline.}
		\label{fig:fig4} 
	\end{figure}
	
	\begin{figure*}
		\centering
		\includegraphics[width=\textwidth]{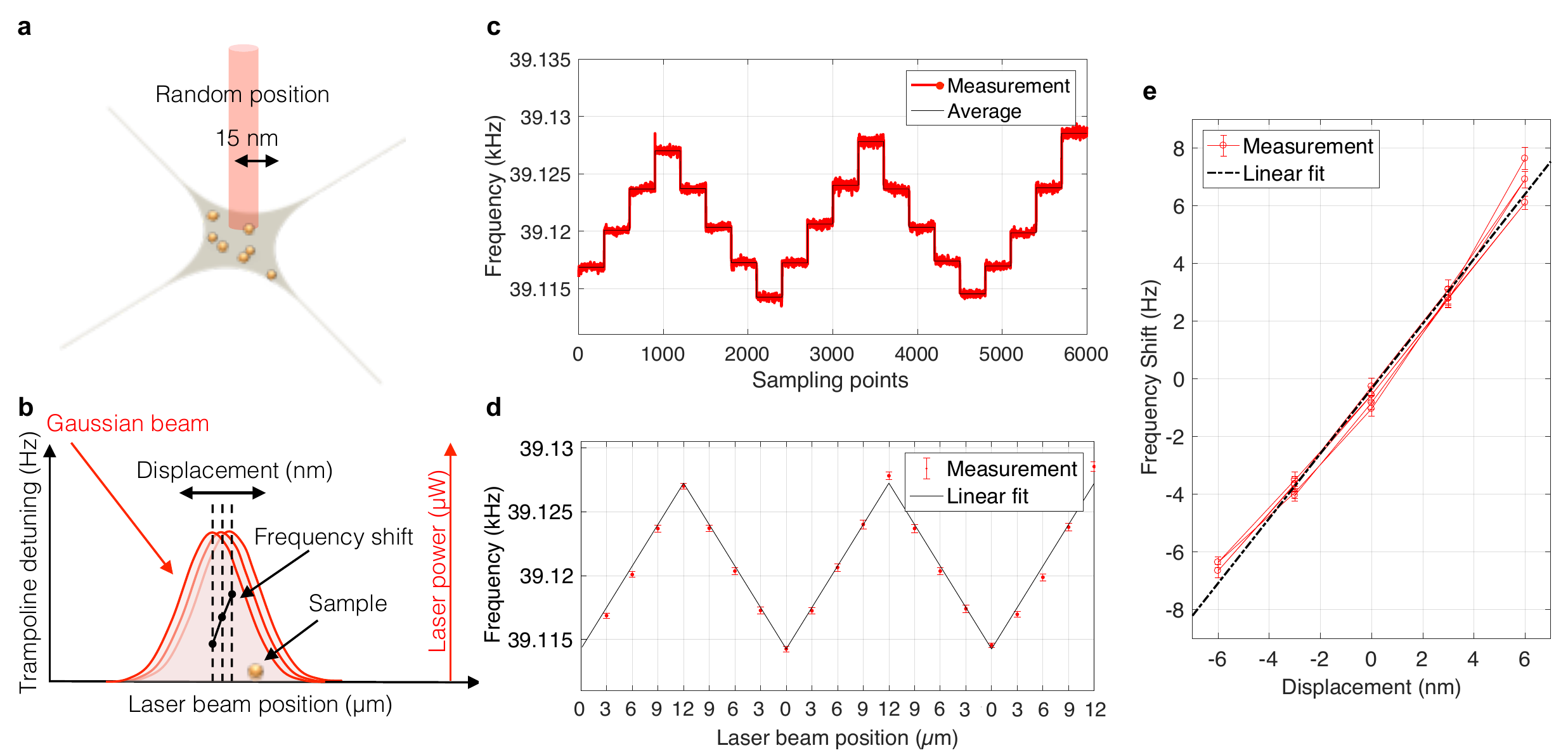} 
		\caption{(a) Schematic of the 15-nm short range fine and repetitive scanning of the single nanoparticle absorber. (b) Schematics of the partly linear behavior of frequency shift within a small displacement region. (c) Raw frequency signal from the PLL, with avergae frequency of each step marked as black solid line. (d) The average and standard deviation of the frequency for each step. The trajectory of the beam movement is marked with black line. (e) The subtracted frequency shift with respect to displacement, with a linear fit to identify responsivity. }
		\label{fig:fig5} 
	\end{figure*}

	\section{Results and Discussion}
	\subsection{Theoretical Model}
	The laser spot displacement $\Delta x$ is measured in terms of the measured relative frequency shift $\delta f = \Delta f/ f_0$  of the trampoline resonator
	\begin{equation}
	\Delta x = \delta R ^{-1} \delta f, 
	\label{eq:resolution}
	\end{equation}
	with the relative responsivity $\delta R$ of the nanoelectromechanical resonator. 
	
	
	The displacement responsivity $\delta R$ of the nanomechanical resonator is defined as its  relative frequency change per laser spot displacement. Since the laser displacement and the resonant frequency change are connected by the change in absorbed laser power $P$, $\delta R$ can be written as
	\begin{equation}\label{eq:dR}
	\delta R = \frac{1}{f_0} \frac{df}{dx} =  \underbrace{\left(\frac{1}{f_0} \frac{df}{dP}\right)}_{\textstyle\delta R_p} \underbrace{\left(\frac{dP}{dx}\right)}_{\textstyle R_x},
	\end{equation}
	with the well-investigated relative power responsivity $\delta R_p$ \cite{schmid2016fundamentals,chien2018} and the beam displacement related power responsivity $R_x$. 
	
	$\delta R_p$ can be modelled  as the responsivity of a nanomechanical cross resonator, which represents the most extreme case of a trampoline that features no central area. A cross resonators has half the responsivity of a nanomechanical string resonator and can be written as \cite{schmid2016fundamentals}
	\begin{equation}\label{eq:dRp}
	\delta R_{p,max} = -\frac{1}{32} \frac{\alpha E}{\kappa \sigma} \frac{L}{wh},
	\end{equation}
	where $\alpha$ is the thermal expansion coefficient, $E$ is the Young's modulus, $\kappa$ is the thermal conductivity, $\sigma$ is the tensile stress, $L$ is the tether length of the cross, and $A$ is the cross-section of the cross resonator tethers. 
	
	Since the dimension of the nanoparticle is around 10\% of the laser beam diameter, the nanoparticle antenna can be approximated as a point absorber with an absorption cross-section $\sigma_{abs}$. Then, the power absorbed by the nanoparticle ($P$) at a position $x$, with respect to a Gaussian beam center, under different beam position, which is the convolution of a point source and a Gaussian beam profile, can be expressed as
	\begin{equation}
	P(x) = \sigma_{abs} \cdot I(x) = \sigma_{abs}\cdot \frac{2 P_0}{\pi w_0^2}  \exp\left(-\frac{2 x^2}{w_0^2}\right),
	\end{equation}
	with the beam radius $w_0$ and the laser power $P_0$, and $I(x)$ as the laser irradiance. According to Equation~(\ref{eq:dR}), $R_x$ can then be obtained by taking the first derivative of $P(x)$ with respect to $x$.
	And since the gradient of a Gaussian beam profile is maximum at half of the beam radius $x = \pm \frac{w_0}{2}$, the maximum $ R_x$ value becomes
	\begin{equation}\label{eq:Rx}
	\delta R_{x,max} = \frac{dP}{dx} \bigg\rvert_{x = \pm \frac{w_0}{2}} \approx 0.8 \, \sigma_{abs}\frac{P_0 }{w_0^3}.
	\end{equation}
	
	Combining Equations (\ref{eq:dRp}) and (\ref{eq:Rx}), an upper limit for the relative displacement responsivity (\ref{eq:dR}) becomes
	\begin{equation}
	\delta R_{max} \approx -0.024 \left(\frac{\alpha E}{\kappa \sigma}\frac{L}{A}\right) \left(\sigma_{abs}\frac{P_0}{w_0^3}\right).
	\label{master}
	\end{equation}
	As indicated by Equation~(\ref{master}), first, the displacement responsivity limit is dependent on the resonator's intrinsic material properties and geometry, and second, on the laser beam profile. In practice this means that for a given resonator material, the NEMS resonator geometry should produce a maximum thermal isolation. And the laser spot should be as small as possible with maximum power.
	
	\subsection{Displacement Responsivity and Resolution}
	
	As has been shown in the derivation of Equation~\ref{eq:Rx}, the displacement responsivity is dependent on the relative position of an absorbing particle with respect to the laser spot, following the first derivative of a Gaussian function. To examine this, a single Au nanoparticle that is well-isolated from other particles in the center area of the trampoline resonator is selected, and a \SI{3x 3}{\micro\meter} area scan  is done with a step size of \SI{80}{\nano\meter} and dwell period of approximately \SI{300}{\milli\second} for each step, as schematically shown in Figure~\ref{fig:fig3}a. As mentioned, the formation of the Gaussian frequency shift profile is a result of the convolution between a single point absorber and the Gaussian power profile of the laser, as illustrated in Figure~\ref{fig:fig3}b. The raw frequency signal for a scan across the single Au particle is plotted in Figure~\ref{fig:fig3}c. The rise time of the NEMS-PSD can be extracted by fitting the step frequency response with a first order exponential function, as shown in the inset of Figure~\ref{fig:fig3}c, yielding a 10\% to 90\% rise time of \SI{53}{\milli\second}. Therefore, a delay time of around \SI{50}{\milli\second} is set for each scan point. More details on the calculation of the rise time is provided in the supplementary information.
	
	Figure~\ref{fig:fig3}d shows the NEMS-PSD frequency shift for a scan across the center of the nanoparticle, as indicated by the white line in the two-dimensional frequency shift mapping in the inset of Figure~\ref{fig:fig3}d. The data points are fitted with a Gaussian function. The extracted full width at half-maximum (FWHM) of the Gaussian fit is \SI{1.1}{\micro\meter}, corresponding to a beam radius of \SI{0.93}{\micro\meter}, which is close to the nominal FWHM of the laser objective (NA = 0.55) of around \SI{0.9}{\micro\meter} (beam radius of \SI{0.75}{\micro\meter}). For the \SI{200}{\nano\meter} Au nanoparticle an absorption cross-section of $\sigma_{abs}=$ \SI{9e-14}{\meter\squared} for a wavelength of \SI{633}{\nano\meter} can be calculated from Mie theory \cite{myroshnychenko2008modelling,chien2018}. With an input laser power of $P_0=$ \SI{85}{\micro\watt}, the expected peak frequency shift is calculated to be approximately \SI{720}{\hertz}, based on the finite element method (FEM) simulations, which fits with the peak frequency shift of \SI{750}{\hertz} in the measurement quite well.
	
	Figure~\ref{fig:fig3}e shows the displacement responsivity, represented by the first derivative of the measured frequency shift profile shown in Figure~\ref{fig:fig3}d.  A maximum responsivity of approximately \SI{1400}{\hertz\per\micro\meter} is reached at half of the beam waist of the Gaussian profile. As expected from the theoretical model, the responsivity is maximal for a particle position at half of the beam radius.
	
	Figure~\ref{fig:fig3}f shows the  Allan deviation for different PLL target bandwidths.
	It can be seen that the thermal drift, represented in the positive slope for the large integration times is consistent for all target bandwidths. In contrast, the negative slopes for short integration times varies for specific target bandwidths, which is because the inductive readout is not limited by thermomechanical noise \cite{demir2019}. From Figure~\ref{fig:fig3}f it can be seen that a target bandwidth of \SI{800}{\hertz} and integration time of \SI{250}{\milli\second} results in a frequency noise of about \SI{250}{\milli\hertz}. This  matches with the standard error of around \SI{253}{\milli\hertz} calculated from the integrated data for individual scan points shown in Figure~\ref{fig:fig3}d (the calculations are presented in more detail in the supplementary information). Together with the previously extracted responsivity of \SI{1400}{\hertz\per\micro\meter}, a position resolution of \SI{105}{\pico\meter\per\sqrt{\hertz}} can be calculated from the current measurement according to Equation~(\ref{eq:resolution}). Since smaller target bandwidths yield lower frequency noise, it's possible to improve the position resolution with the trade-off of longer minimal integration times. E.g. using a target bandwidth of \SI{10}{\hertz} results in a minimum frequency noise of \SI{8}{\milli\hertz} for an integration time of \SI{450}{\milli\second}, which ultimately results in a position resolution of \SI{4}{\pico\meter\per\sqrt{\hertz}}. 
	
	\subsection{Optimization of Trampoline Geometry}
	
	As discussed previously in Equation~\ref{eq:dR}, the maximum responsivity depends on the design of the trampoline geometry. Figure~\ref{fig:fig4} presents the study of trampoline resonators with various sizes of the center area, both measured and simulated using FEM. The window size is kept constant  at \SI{1}{\milli\meter} for all designs. The width of the trampoline tethers were kept constant at  \SI{5}{\micro\meter} and a silicon nitride thickness of \SI{50}{\nano\meter}. The curvatures of all trampoline geometries are optimized for an even stress distribution. Single \SI{200}{\nano\meter} gold nanoparticles are used as absorber for all measurements, and the maximum displacement responsivity is extracted as shown in Figure~\ref{fig:fig3}e. The measured responsivities share a similar trend with the FEM simulation, with decreasing responsivity for increasing center area from type A to type E trampoline. This can be explained by the resulting lower temperature profile for trampolines with larger center area, as shown in more detail in the supplementary Figure~\ref{fig:figS5}. The most extreme design of type A trampoline has a responsivity close to the theoretical limit according to Equation~\ref{master} of around \SI{2100}{\hertz\per\micro\meter}, representing a trampoline with no center area.
	
	However, the small center area makes it challenging to distribute nanoparticles on the surface by means of spin-coating. Type C trampolines have a large enough center area and show a high responsivity, only 10\% less compared to the most responsive type A trampoline. Therefore, type C trampolines were the devices of choice  used for the present study. Furthermore, since the nanoparticles are distributed randomly on the trampoline resonator, the position-dependency of the power responsivity is worth discussing. As shown in the finite element simulation in the inset of Figure~\ref{fig:fig4}, the power responsivity remains constant over the entire center area. It's not until the tethers of the trampoline resonator are reached that the responsivity starts to drop. Hence, the displacement measurement can be operated optimally over the whole center area, which is a pre-condition for the displacement sensing with type C trampoline resonators. 
	
	\subsection{Repetitive Displacement Measurements}
	
	After the characterization of the displacement responsivity and the optimization of the trampoline geometry, a repetitive displacement measurement within a small range of \SI{15}{\nano\meter} and small step size of \SI{3}{\nano\meter} is performed. Therefore, the laser spot was placed randomly on the center of the trampoline in order to demonstrate the possibility to operate the NEMS-PSD without any fine alignment and calibration process demonstrating the practicability.  Furthermore, the repeatability and long-term stability of displacement sensing is studied, as illustrated in Figure~\ref{fig:fig5}a. The integration time for each step remains \SI{250}{\milli\second}. As in the schematics of Figure~\ref{fig:fig5}b, the frequency shift is no longer a Gaussian profile, instead, an approximated linear relation between the frequency shift of the trampoline resonator and the beam displacement with an almost constant displacement responsivity can be expected for such a small scan range. The frequency signal from the PLL is plotted in Figure~\ref{fig:fig5}c, where the individual displacement steps of \SI{3}{\nano\meter} can be clearly identified. The averaged frequency and standard deviation of each displacement step is then calculated and plotted in Figure~\ref{fig:fig5}d with respect to the laser beam position. For generalization, a reference position is then defined as the center point of the repetitive movement, and the frequency shift is then subtracted according to this reference frequency, as shown in Figure~\ref{fig:fig5}e. A linear fit is subsequently performed to extract the displacement responsivity within this range. The averaged displacement responsivity obtained from the fit is \SI{1.15}{\hertz\per\nano\meter}. The target bandwidth of the PLL is reduced to \SI{200}{\hertz} in these measurements, resulting in a frequency noise of around \SI{0.025}{\hertz}. This gives a position resolution of \SI{12}{\pico\meter\per\sqrt{\hertz}}, which is slightly worse than previous values, which were obtained for an optimized particle/beam position. With this measurement, NEMS-PSD demonstrates the feasibility of operation without fine optical alignment on a random absorber on the trampoline resonator.
	
	
	The repeatability after 5 cycles of operation is within a maximum deviation of \SI{2}{\nano\meter}. This deviation could also partly result from the accuracy of the nanopositioning stage that controls the position of the beam. With a drift below \SI{500}{\nano\meter} within 4 hours, as discussed in the supplementary information, the long-term stability of NEMS-PSD is of the same order as of photodiode-based PSDs with \SIrange{0.1}{1}{\micro\meter} \cite{haddad2008gaussian,makynen2000position}. The drift could be partly contributed by the whole optical system including sample mounting. This high repeatability and long-term stability of NEMS-PSD can result from the localized absorption of nanoparticle, making the system less susceptible to background scattering and interference.
	
	
	\section{Conclusions}
	We presented a NEMS-PSD based on silicon nitride trampoline resonators with integrated electrodynamic readout and actuation. We demonstrated a sensitivity of \SI{4}{\pico\meter\per\sqrt{\hertz}} with the potential of further optimization by using e.g. silicon nitride with lower stress. The NEMS-PSD demonstrated a repeatability of approximately \SI{2}{\nano\meter} after 5 cycles of operation and showed a long-term stability better than \SI{500}{\nano\meter} in 4 hours. This position-sensitive detector design overcomes the issue of non-uniformity of multiple segments by measuring the direct absorption instead of differential current from photodiodes, which improves the sensitivity greatly, and also requires minimum signal processing effort. Due to the localized nanoparticle absorber, the parasitic effect from the ambient is also minimized, which enables better long-term stability. It is also compatible with small beam diameter and even irregular beams, since the artifact could be easily calibrated with a scan before operation to identify the beam profile. The presented NEMS-PSD is promising to provide a sensitive alternative to existing PSDs and could bring advances to a great variety of research and application fields.
	

	\section{Methods}
	
	\subsection{Sample Fabrication}
	The samples are fabricated with a bulk micromaching process. A silicon wafer with a thickness of \SI{370}{\micro\meter} is deposited with 50 nm silicon-rich silicon nitride (SiN) with low pressure chemical vapor deposition (LPCVD). The prestress is approximately \SI{150}{\mega\pascal}. 190~nm thick gold electrodes together with a 10~nm chrome adhesion layer for magnetomotive transduction is first defined with photolithography on the front side of the SiN wafer, and the SiN trampoline structure is then defined with another step of photolithography after lift-off. The excess SiN is then removed with reactive ion etching (RIE) and protected with a layer of photoresist. A window is defined from the back side and etched with KOH to release the trampoline resonator. Reactant-free gold nanoparticles with a diameter of 200 nm in 0.1 mM PBS stabilized suspension solution (SigmaAldrich) are first diluted in Micropur deionized water (\SI{18}{\mega\ohm}, Milli-Q) at a ratio of 1:100 at room temperature, and then spin-coated on the trampoline resonator at 2000 rpm.
	
	\subsection{Finite Element Method Simulation}
	The finite element simulations are done with the thermal stress module of COMSOL multiphysics, including first the simulation of the temperature field of a point heat source and subsequently the stress distribution and the eigenfrequency. The responsivity could be extracted by simulating the eigenfrequency at different powers of the point heat source. The thermal expansion coefficient ($\alpha$) used in the simulation is \SI{2.2e-6}{\per\kelvin}, the Young's modulus (E) \SI{250}{\giga\pascal}, thermal conductivity ($\kappa$) \SI{3}{\watt\per\meter\per\kelvin}, and prestress ($\sigma$) of \SI{150}{\mega\pascal}. All the constants are also consistent with the ones used for theoretical calculations.
	
	\subsection{Measurement Setup}
	The optical setup is shown in detail in Figure~\ref{fig:figS2}. In this experiment, a diode laser with 633 nm wavelength (Toptica TopMode) is used. The beam passes through a beam expander and the power is reduced to approximately \SI{85}{\micro\watt} before the vacuum chamber with a linear polarizer. A 50 times objective (0.55 N.A.; Mitutoyo) is mounted on the nanopositioning stage (PiMars, Physikinstrumente) for control of the beam position. All the measurements are done under a vacuum of \SI{1e-3}{\milli\bar}. The magnetomotive transduction is done with an enhanced Halbach array with the layout shown in Figure~\ref{fig:figS1}. The magnetic field in the center 5 mm region is measured to be above \SI{1}{\tesla}. The electrical signal from the trampoline resonator is first amplified with a home-built low-noise pre-amplifier (LT1028, Analog Devices) with a gain of 200, and fed to the lock-in amplifier with a phase-locked loop (HF2LI, Zurich Instrument), with its output driving the actuation.
	
	\begin{acknowledgments}
		We gratefully acknowledge the assistance of Sophia Ewert and Patrick Meyer with the sample fabrication and preparation, and the useful discussions with Markus Piller and Hendrik K\"ahler. This work is supported by the European Research Council under the European Unions Horizon 2020 research and innovation program (Grant Agreement-716087-PLASMECS). 
	\end{acknowledgments}

	\bibliography{displacement_sensing.bib}

\begin{thebibliography}{23}%
\makeatletter
\providecommand \@ifxundefined [1]{%
 \@ifx{#1\undefined}
}%
\providecommand \@ifnum [1]{%
 \ifnum #1\expandafter \@firstoftwo
 \else \expandafter \@secondoftwo
 \fi
}%
\providecommand \@ifx [1]{%
 \ifx #1\expandafter \@firstoftwo
 \else \expandafter \@secondoftwo
 \fi
}%
\providecommand \natexlab [1]{#1}%
\providecommand \enquote  [1]{``#1''}%
\providecommand \bibnamefont  [1]{#1}%
\providecommand \bibfnamefont [1]{#1}%
\providecommand \citenamefont [1]{#1}%
\providecommand \href@noop [0]{\@secondoftwo}%
\providecommand \href [0]{\begingroup \@sanitize@url \@href}%
\providecommand \@href[1]{\@@startlink{#1}\@@href}%
\providecommand \@@href[1]{\endgroup#1\@@endlink}%
\providecommand \@sanitize@url [0]{\catcode `\\12\catcode `\$12\catcode
  `\&12\catcode `\#12\catcode `\^12\catcode `\_12\catcode `\%12\relax}%
\providecommand \@@startlink[1]{}%
\providecommand \@@endlink[0]{}%
\providecommand \url  [0]{\begingroup\@sanitize@url \@url }%
\providecommand \@url [1]{\endgroup\@href {#1}{\urlprefix }}%
\providecommand \urlprefix  [0]{URL }%
\providecommand \Eprint [0]{\href }%
\providecommand \doibase [0]{http://dx.doi.org/}%
\providecommand \selectlanguage [0]{\@gobble}%
\providecommand \bibinfo  [0]{\@secondoftwo}%
\providecommand \bibfield  [0]{\@secondoftwo}%
\providecommand \translation [1]{[#1]}%
\providecommand \BibitemOpen [0]{}%
\providecommand \bibitemStop [0]{}%
\providecommand \bibitemNoStop [0]{.\EOS\space}%
\providecommand \EOS [0]{\spacefactor3000\relax}%
\providecommand \BibitemShut  [1]{\csname bibitem#1\endcsname}%
\let\auto@bib@innerbib\@empty
\bibitem [{\citenamefont {Dufr{\^e}ne}\ \emph {et~al.}(2017)\citenamefont
  {Dufr{\^e}ne}, \citenamefont {Ando}, \citenamefont {Garcia}, \citenamefont
  {Alsteens}, \citenamefont {Martinez-Martin}, \citenamefont {Engel},
  \citenamefont {Gerber},\ and\ \citenamefont
  {M{\"u}ller}}]{dufrene2017imaging}%
  \BibitemOpen
  \bibfield  {author} {\bibinfo {author} {\bibfnamefont {Y.~F.}\ \bibnamefont
  {Dufr{\^e}ne}}, \bibinfo {author} {\bibfnamefont {T.}~\bibnamefont {Ando}},
  \bibinfo {author} {\bibfnamefont {R.}~\bibnamefont {Garcia}}, \bibinfo
  {author} {\bibfnamefont {D.}~\bibnamefont {Alsteens}}, \bibinfo {author}
  {\bibfnamefont {D.}~\bibnamefont {Martinez-Martin}}, \bibinfo {author}
  {\bibfnamefont {A.}~\bibnamefont {Engel}}, \bibinfo {author} {\bibfnamefont
  {C.}~\bibnamefont {Gerber}}, \ and\ \bibinfo {author} {\bibfnamefont {D.~J.}\
  \bibnamefont {M{\"u}ller}},\ }\href@noop {} {\bibfield  {journal} {\bibinfo
  {journal} {Nature nanotechnology}\ }\textbf {\bibinfo {volume} {12}},\
  \bibinfo {pages} {295} (\bibinfo {year} {2017})}\BibitemShut {NoStop}%
\bibitem [{\citenamefont {Krieg}\ \emph {et~al.}(2019)\citenamefont {Krieg},
  \citenamefont {Fl{\"a}schner}, \citenamefont {Alsteens}, \citenamefont
  {Gaub}, \citenamefont {Roos}, \citenamefont {Wuite}, \citenamefont {Gaub},
  \citenamefont {Gerber}, \citenamefont {Dufr{\^e}ne},\ and\ \citenamefont
  {M{\"u}ller}}]{krieg2019atomic}%
  \BibitemOpen
  \bibfield  {author} {\bibinfo {author} {\bibfnamefont {M.}~\bibnamefont
  {Krieg}}, \bibinfo {author} {\bibfnamefont {G.}~\bibnamefont
  {Fl{\"a}schner}}, \bibinfo {author} {\bibfnamefont {D.}~\bibnamefont
  {Alsteens}}, \bibinfo {author} {\bibfnamefont {B.~M.}\ \bibnamefont {Gaub}},
  \bibinfo {author} {\bibfnamefont {W.~H.}\ \bibnamefont {Roos}}, \bibinfo
  {author} {\bibfnamefont {G.~J.}\ \bibnamefont {Wuite}}, \bibinfo {author}
  {\bibfnamefont {H.~E.}\ \bibnamefont {Gaub}}, \bibinfo {author}
  {\bibfnamefont {C.}~\bibnamefont {Gerber}}, \bibinfo {author} {\bibfnamefont
  {Y.~F.}\ \bibnamefont {Dufr{\^e}ne}}, \ and\ \bibinfo {author} {\bibfnamefont
  {D.~J.}\ \bibnamefont {M{\"u}ller}},\ }\href@noop {} {\bibfield  {journal}
  {\bibinfo  {journal} {Nature Reviews Physics}\ }\textbf {\bibinfo {volume}
  {1}},\ \bibinfo {pages} {41} (\bibinfo {year} {2019})}\BibitemShut {NoStop}%
\bibitem [{\citenamefont {Knobel}\ and\ \citenamefont
  {Cleland}(2003)}]{knobel2003nanometre}%
  \BibitemOpen
  \bibfield  {author} {\bibinfo {author} {\bibfnamefont {R.~G.}\ \bibnamefont
  {Knobel}}\ and\ \bibinfo {author} {\bibfnamefont {A.~N.}\ \bibnamefont
  {Cleland}},\ }\href@noop {} {\bibfield  {journal} {\bibinfo  {journal}
  {Nature}\ }\textbf {\bibinfo {volume} {424}},\ \bibinfo {pages} {291}
  (\bibinfo {year} {2003})}\BibitemShut {NoStop}%
\bibitem [{\citenamefont {Naik}\ \emph {et~al.}(2009)\citenamefont {Naik},
  \citenamefont {Hanay}, \citenamefont {Hiebert}, \citenamefont {Feng},\ and\
  \citenamefont {Roukes}}]{naik2009towards}%
  \BibitemOpen
  \bibfield  {author} {\bibinfo {author} {\bibfnamefont {A.~K.}\ \bibnamefont
  {Naik}}, \bibinfo {author} {\bibfnamefont {M.}~\bibnamefont {Hanay}},
  \bibinfo {author} {\bibfnamefont {W.}~\bibnamefont {Hiebert}}, \bibinfo
  {author} {\bibfnamefont {X.}~\bibnamefont {Feng}}, \ and\ \bibinfo {author}
  {\bibfnamefont {M.~L.}\ \bibnamefont {Roukes}},\ }\href@noop {} {\bibfield
  {journal} {\bibinfo  {journal} {Nature nanotechnology}\ }\textbf {\bibinfo
  {volume} {4}},\ \bibinfo {pages} {445} (\bibinfo {year} {2009})}\BibitemShut
  {NoStop}%
\bibitem [{\citenamefont {Hanay}\ \emph {et~al.}(2012)\citenamefont {Hanay},
  \citenamefont {Kelber}, \citenamefont {Naik}, \citenamefont {Chi},
  \citenamefont {Hentz}, \citenamefont {Bullard}, \citenamefont {Colinet},
  \citenamefont {Duraffourg},\ and\ \citenamefont {Roukes}}]{hanay2012single}%
  \BibitemOpen
  \bibfield  {author} {\bibinfo {author} {\bibfnamefont {M.~S.}\ \bibnamefont
  {Hanay}}, \bibinfo {author} {\bibfnamefont {S.}~\bibnamefont {Kelber}},
  \bibinfo {author} {\bibfnamefont {A.}~\bibnamefont {Naik}}, \bibinfo {author}
  {\bibfnamefont {D.}~\bibnamefont {Chi}}, \bibinfo {author} {\bibfnamefont
  {S.}~\bibnamefont {Hentz}}, \bibinfo {author} {\bibfnamefont
  {E.}~\bibnamefont {Bullard}}, \bibinfo {author} {\bibfnamefont
  {E.}~\bibnamefont {Colinet}}, \bibinfo {author} {\bibfnamefont
  {L.}~\bibnamefont {Duraffourg}}, \ and\ \bibinfo {author} {\bibfnamefont
  {M.}~\bibnamefont {Roukes}},\ }\href@noop {} {\bibfield  {journal} {\bibinfo
  {journal} {Nature nanotechnology}\ }\textbf {\bibinfo {volume} {7}},\
  \bibinfo {pages} {602} (\bibinfo {year} {2012})}\BibitemShut {NoStop}%
\bibitem [{\citenamefont {Sage}\ \emph {et~al.}(2015)\citenamefont {Sage},
  \citenamefont {Brenac}, \citenamefont {Alava}, \citenamefont {Morel},
  \citenamefont {Dupr{\'e}}, \citenamefont {Hanay}, \citenamefont {Roukes},
  \citenamefont {Duraffourg}, \citenamefont {Masselon},\ and\ \citenamefont
  {Hentz}}]{sage2015neutral}%
  \BibitemOpen
  \bibfield  {author} {\bibinfo {author} {\bibfnamefont {E.}~\bibnamefont
  {Sage}}, \bibinfo {author} {\bibfnamefont {A.}~\bibnamefont {Brenac}},
  \bibinfo {author} {\bibfnamefont {T.}~\bibnamefont {Alava}}, \bibinfo
  {author} {\bibfnamefont {R.}~\bibnamefont {Morel}}, \bibinfo {author}
  {\bibfnamefont {C.}~\bibnamefont {Dupr{\'e}}}, \bibinfo {author}
  {\bibfnamefont {M.~S.}\ \bibnamefont {Hanay}}, \bibinfo {author}
  {\bibfnamefont {M.~L.}\ \bibnamefont {Roukes}}, \bibinfo {author}
  {\bibfnamefont {L.}~\bibnamefont {Duraffourg}}, \bibinfo {author}
  {\bibfnamefont {C.}~\bibnamefont {Masselon}}, \ and\ \bibinfo {author}
  {\bibfnamefont {S.}~\bibnamefont {Hentz}},\ }\href@noop {} {\bibfield
  {journal} {\bibinfo  {journal} {Nature communications}\ }\textbf {\bibinfo
  {volume} {6}},\ \bibinfo {pages} {1} (\bibinfo {year} {2015})}\BibitemShut
  {NoStop}%
\bibitem [{\citenamefont {Biercuk}\ \emph {et~al.}(2010)\citenamefont
  {Biercuk}, \citenamefont {Uys}, \citenamefont {Britton}, \citenamefont
  {VanDevender},\ and\ \citenamefont {Bollinger}}]{biercuk2010}%
  \BibitemOpen
  \bibfield  {author} {\bibinfo {author} {\bibfnamefont {M.~J.}\ \bibnamefont
  {Biercuk}}, \bibinfo {author} {\bibfnamefont {H.}~\bibnamefont {Uys}},
  \bibinfo {author} {\bibfnamefont {J.~W.}\ \bibnamefont {Britton}}, \bibinfo
  {author} {\bibfnamefont {A.~P.}\ \bibnamefont {VanDevender}}, \ and\ \bibinfo
  {author} {\bibfnamefont {J.~J.}\ \bibnamefont {Bollinger}},\ }\href@noop {}
  {\bibfield  {journal} {\bibinfo  {journal} {Nature nanotechnology}\ }\textbf
  {\bibinfo {volume} {5}},\ \bibinfo {pages} {646} (\bibinfo {year}
  {2010})}\BibitemShut {NoStop}%
\bibitem [{\citenamefont {Grinolds}\ \emph {et~al.}(2013)\citenamefont
  {Grinolds}, \citenamefont {Hong}, \citenamefont {Maletinsky}, \citenamefont
  {Luan}, \citenamefont {Lukin}, \citenamefont {Walsworth},\ and\ \citenamefont
  {Yacoby}}]{grinolds2013nanoscale}%
  \BibitemOpen
  \bibfield  {author} {\bibinfo {author} {\bibfnamefont {M.~S.}\ \bibnamefont
  {Grinolds}}, \bibinfo {author} {\bibfnamefont {S.}~\bibnamefont {Hong}},
  \bibinfo {author} {\bibfnamefont {P.}~\bibnamefont {Maletinsky}}, \bibinfo
  {author} {\bibfnamefont {L.}~\bibnamefont {Luan}}, \bibinfo {author}
  {\bibfnamefont {M.~D.}\ \bibnamefont {Lukin}}, \bibinfo {author}
  {\bibfnamefont {R.~L.}\ \bibnamefont {Walsworth}}, \ and\ \bibinfo {author}
  {\bibfnamefont {A.}~\bibnamefont {Yacoby}},\ }\href@noop {} {\bibfield
  {journal} {\bibinfo  {journal} {Nature Physics}\ }\textbf {\bibinfo {volume}
  {9}},\ \bibinfo {pages} {215} (\bibinfo {year} {2013})}\BibitemShut {NoStop}%
\bibitem [{\citenamefont {M{\"a}kynen}(2000)}]{makynen2000position}%
  \BibitemOpen
  \bibfield  {author} {\bibinfo {author} {\bibfnamefont {A.}~\bibnamefont
  {M{\"a}kynen}},\ }\href@noop {} {\emph {\bibinfo {title} {Position-sensitive
  devices and sensor systems for optical tracking and displacement sensing
  applications}}}\ (\bibinfo  {publisher} {Oulun yliopisto},\ \bibinfo {year}
  {2000})\BibitemShut {NoStop}%
\bibitem [{\citenamefont {Andersson}(2008)}]{andersson2008position}%
  \BibitemOpen
  \bibfield  {author} {\bibinfo {author} {\bibfnamefont {H.}~\bibnamefont
  {Andersson}},\ }\emph {\bibinfo {title} {Position sensitive detectors: device
  technology and applications in spectroscopy}},\ \href@noop {} {Ph.D.
  thesis},\ \bibinfo  {school} {Institutionen f{\"o}r informationsteknologi och
  medier} (\bibinfo {year} {2008})\BibitemShut {NoStop}%
\bibitem [{\citenamefont {Azaryan}\ \emph {et~al.}(2019)\citenamefont
  {Azaryan}, \citenamefont {Budagov}, \citenamefont {Lyablin}, \citenamefont
  {Pluzhnikov}, \citenamefont {Di~Girolamo}, \citenamefont {Gayde},\ and\
  \citenamefont {Mergelkuhl}}]{azaryan2019position}%
  \BibitemOpen
  \bibfield  {author} {\bibinfo {author} {\bibfnamefont {N.~S.}\ \bibnamefont
  {Azaryan}}, \bibinfo {author} {\bibfnamefont {J.~A.}\ \bibnamefont
  {Budagov}}, \bibinfo {author} {\bibfnamefont {M.}~\bibnamefont {Lyablin}},
  \bibinfo {author} {\bibfnamefont {A.~A.}\ \bibnamefont {Pluzhnikov}},
  \bibinfo {author} {\bibfnamefont {B.}~\bibnamefont {Di~Girolamo}}, \bibinfo
  {author} {\bibfnamefont {J.-C.}\ \bibnamefont {Gayde}}, \ and\ \bibinfo
  {author} {\bibfnamefont {D.}~\bibnamefont {Mergelkuhl}},\ }\href@noop {}
  {\bibfield  {journal} {\bibinfo  {journal} {Physics of Particles and Nuclei
  Letters}\ }\textbf {\bibinfo {volume} {16}},\ \bibinfo {pages} {354}
  (\bibinfo {year} {2019})}\BibitemShut {NoStop}%
\bibitem [{\citenamefont {Haddad}\ \emph {et~al.}(2008)\citenamefont {Haddad},
  \citenamefont {Juncar}, \citenamefont {Geneves},\ and\ \citenamefont
  {Wakim}}]{haddad2008gaussian}%
  \BibitemOpen
  \bibfield  {author} {\bibinfo {author} {\bibfnamefont {D.}~\bibnamefont
  {Haddad}}, \bibinfo {author} {\bibfnamefont {P.}~\bibnamefont {Juncar}},
  \bibinfo {author} {\bibfnamefont {G.}~\bibnamefont {Geneves}}, \ and\
  \bibinfo {author} {\bibfnamefont {M.}~\bibnamefont {Wakim}},\ }\href@noop {}
  {\bibfield  {journal} {\bibinfo  {journal} {ieee Transactions on
  Instrumentation and Measurement}\ }\textbf {\bibinfo {volume} {58}},\
  \bibinfo {pages} {1003} (\bibinfo {year} {2008})}\BibitemShut {NoStop}%
\bibitem [{\citenamefont {Bag}\ \emph {et~al.}(2019)\citenamefont {Bag},
  \citenamefont {Neugebauer}, \citenamefont {Mick}, \citenamefont
  {Christiansen}, \citenamefont {Schulz},\ and\ \citenamefont
  {Banzer}}]{bag2019}%
  \BibitemOpen
  \bibfield  {author} {\bibinfo {author} {\bibfnamefont {A.}~\bibnamefont
  {Bag}}, \bibinfo {author} {\bibfnamefont {M.}~\bibnamefont {Neugebauer}},
  \bibinfo {author} {\bibfnamefont {U.}~\bibnamefont {Mick}}, \bibinfo {author}
  {\bibfnamefont {S.}~\bibnamefont {Christiansen}}, \bibinfo {author}
  {\bibfnamefont {S.~A.}\ \bibnamefont {Schulz}}, \ and\ \bibinfo {author}
  {\bibfnamefont {P.}~\bibnamefont {Banzer}},\ }\href@noop {} {\bibfield
  {journal} {\bibinfo  {journal} {arXiv preprint arXiv:1909.04478}\ } (\bibinfo
  {year} {2019})}\BibitemShut {NoStop}%
\bibitem [{\citenamefont {Piller}\ \emph {et~al.}(2019)\citenamefont {Piller},
  \citenamefont {Luhmann}, \citenamefont {Chien},\ and\ \citenamefont
  {Schmid}}]{piller2019}%
  \BibitemOpen
  \bibfield  {author} {\bibinfo {author} {\bibfnamefont {M.}~\bibnamefont
  {Piller}}, \bibinfo {author} {\bibfnamefont {N.}~\bibnamefont {Luhmann}},
  \bibinfo {author} {\bibfnamefont {M.-H.}\ \bibnamefont {Chien}}, \ and\
  \bibinfo {author} {\bibfnamefont {S.}~\bibnamefont {Schmid}},\ }in\
  \href@noop {} {\emph {\bibinfo {booktitle} {Optical Sensing, Imaging, and
  Photon Counting: From X-Rays to THz 2019}}},\ Vol.\ \bibinfo {volume}
  {11088}\ (\bibinfo {organization} {International Society for Optics and
  Photonics},\ \bibinfo {year} {2019})\ p.\ \bibinfo {pages}
  {1108802}\BibitemShut {NoStop}%
\bibitem [{\citenamefont {Schmid}\ \emph {et~al.}(2014)\citenamefont {Schmid},
  \citenamefont {Wu}, \citenamefont {Larsen}, \citenamefont {Rindzevicius},\
  and\ \citenamefont {Boisen}}]{schmid2014low}%
  \BibitemOpen
  \bibfield  {author} {\bibinfo {author} {\bibfnamefont {S.}~\bibnamefont
  {Schmid}}, \bibinfo {author} {\bibfnamefont {K.}~\bibnamefont {Wu}}, \bibinfo
  {author} {\bibfnamefont {P.~E.}\ \bibnamefont {Larsen}}, \bibinfo {author}
  {\bibfnamefont {T.}~\bibnamefont {Rindzevicius}}, \ and\ \bibinfo {author}
  {\bibfnamefont {A.}~\bibnamefont {Boisen}},\ }\href@noop {} {\bibfield
  {journal} {\bibinfo  {journal} {Nano letters}\ }\textbf {\bibinfo {volume}
  {14}},\ \bibinfo {pages} {2318} (\bibinfo {year} {2014})}\BibitemShut
  {NoStop}%
\bibitem [{\citenamefont {Larsen}\ \emph {et~al.}(2013)\citenamefont {Larsen},
  \citenamefont {Schmid}, \citenamefont {Villanueva},\ and\ \citenamefont
  {Boisen}}]{larsen2013photothermal}%
  \BibitemOpen
  \bibfield  {author} {\bibinfo {author} {\bibfnamefont {T.}~\bibnamefont
  {Larsen}}, \bibinfo {author} {\bibfnamefont {S.}~\bibnamefont {Schmid}},
  \bibinfo {author} {\bibfnamefont {L.~G.}\ \bibnamefont {Villanueva}}, \ and\
  \bibinfo {author} {\bibfnamefont {A.}~\bibnamefont {Boisen}},\ }\href@noop {}
  {\bibfield  {journal} {\bibinfo  {journal} {ACS nano}\ }\textbf {\bibinfo
  {volume} {7}},\ \bibinfo {pages} {6188} (\bibinfo {year} {2013})}\BibitemShut
  {NoStop}%
\bibitem [{\citenamefont {Chien}\ \emph {et~al.}(2018)\citenamefont {Chien},
  \citenamefont {Brameshuber}, \citenamefont {Rossboth}, \citenamefont
  {Sch{\"u}tz},\ and\ \citenamefont {Schmid}}]{chien2018}%
  \BibitemOpen
  \bibfield  {author} {\bibinfo {author} {\bibfnamefont {M.-H.}\ \bibnamefont
  {Chien}}, \bibinfo {author} {\bibfnamefont {M.}~\bibnamefont {Brameshuber}},
  \bibinfo {author} {\bibfnamefont {B.~K.}\ \bibnamefont {Rossboth}}, \bibinfo
  {author} {\bibfnamefont {G.~J.}\ \bibnamefont {Sch{\"u}tz}}, \ and\ \bibinfo
  {author} {\bibfnamefont {S.}~\bibnamefont {Schmid}},\ }\href@noop {}
  {\bibfield  {journal} {\bibinfo  {journal} {Proceedings of the National
  Academy of Sciences}\ }\textbf {\bibinfo {volume} {115}},\ \bibinfo {pages}
  {11150} (\bibinfo {year} {2018})}\BibitemShut {NoStop}%
\bibitem [{\citenamefont {Cleland}\ and\ \citenamefont
  {Roukes}(1996)}]{cleland1996}%
  \BibitemOpen
  \bibfield  {author} {\bibinfo {author} {\bibfnamefont {A.~N.}\ \bibnamefont
  {Cleland}}\ and\ \bibinfo {author} {\bibfnamefont {M.~L.}\ \bibnamefont
  {Roukes}},\ }\href@noop {} {\bibfield  {journal} {\bibinfo  {journal}
  {Applied Physics Letters}\ }\textbf {\bibinfo {volume} {69}},\ \bibinfo
  {pages} {2653} (\bibinfo {year} {1996})}\BibitemShut {NoStop}%
\bibitem [{\citenamefont {Cleland}\ and\ \citenamefont
  {Roukes}(1999)}]{cleland1999}%
  \BibitemOpen
  \bibfield  {author} {\bibinfo {author} {\bibfnamefont {A.}~\bibnamefont
  {Cleland}}\ and\ \bibinfo {author} {\bibfnamefont {M.}~\bibnamefont
  {Roukes}},\ }\href@noop {} {\bibfield  {journal} {\bibinfo  {journal}
  {Sensors and Actuators A: Physical}\ }\textbf {\bibinfo {volume} {72}},\
  \bibinfo {pages} {256} (\bibinfo {year} {1999})}\BibitemShut {NoStop}%
\bibitem [{\citenamefont {Venstra}\ \emph {et~al.}(2009)\citenamefont
  {Venstra}, \citenamefont {Westra}, \citenamefont {Gavan},\ and\ \citenamefont
  {Van~der Zant}}]{venstra2009}%
  \BibitemOpen
  \bibfield  {author} {\bibinfo {author} {\bibfnamefont {W.}~\bibnamefont
  {Venstra}}, \bibinfo {author} {\bibfnamefont {H.}~\bibnamefont {Westra}},
  \bibinfo {author} {\bibfnamefont {K.~B.}\ \bibnamefont {Gavan}}, \ and\
  \bibinfo {author} {\bibfnamefont {H.}~\bibnamefont {Van~der Zant}},\
  }\href@noop {} {\bibfield  {journal} {\bibinfo  {journal} {Applied Physics
  Letters}\ }\textbf {\bibinfo {volume} {95}},\ \bibinfo {pages} {263103}
  (\bibinfo {year} {2009})}\BibitemShut {NoStop}%
\bibitem [{\citenamefont {Schmid}\ \emph {et~al.}(2016)\citenamefont {Schmid},
  \citenamefont {Villanueva},\ and\ \citenamefont
  {Roukes}}]{schmid2016fundamentals}%
  \BibitemOpen
  \bibfield  {author} {\bibinfo {author} {\bibfnamefont {S.}~\bibnamefont
  {Schmid}}, \bibinfo {author} {\bibfnamefont {L.~G.}\ \bibnamefont
  {Villanueva}}, \ and\ \bibinfo {author} {\bibfnamefont {M.~L.}\ \bibnamefont
  {Roukes}},\ }\href@noop {} {\emph {\bibinfo {title} {Fundamentals of
  nanomechanical resonators}}},\ Vol.~\bibinfo {volume} {49}\ (\bibinfo
  {publisher} {Springer},\ \bibinfo {year} {2016})\BibitemShut {NoStop}%
\bibitem [{\citenamefont {Myroshnychenko}\ \emph {et~al.}(2008)\citenamefont
  {Myroshnychenko}, \citenamefont {Rodriguez-Fernandez}, \citenamefont
  {Pastoriza-Santos}, \citenamefont {Funston}, \citenamefont {Novo},
  \citenamefont {Mulvaney}, \citenamefont {Liz-Marz{\'a}n},\ and\ \citenamefont
  {De~Abajo}}]{myroshnychenko2008modelling}%
  \BibitemOpen
  \bibfield  {author} {\bibinfo {author} {\bibfnamefont {V.}~\bibnamefont
  {Myroshnychenko}}, \bibinfo {author} {\bibfnamefont {J.}~\bibnamefont
  {Rodriguez-Fernandez}}, \bibinfo {author} {\bibfnamefont {I.}~\bibnamefont
  {Pastoriza-Santos}}, \bibinfo {author} {\bibfnamefont {A.~M.}\ \bibnamefont
  {Funston}}, \bibinfo {author} {\bibfnamefont {C.}~\bibnamefont {Novo}},
  \bibinfo {author} {\bibfnamefont {P.}~\bibnamefont {Mulvaney}}, \bibinfo
  {author} {\bibfnamefont {L.~M.}\ \bibnamefont {Liz-Marz{\'a}n}}, \ and\
  \bibinfo {author} {\bibfnamefont {F.~J.~G.}\ \bibnamefont {De~Abajo}},\
  }\href@noop {} {\bibfield  {journal} {\bibinfo  {journal} {Chemical Society
  Reviews}\ }\textbf {\bibinfo {volume} {37}},\ \bibinfo {pages} {1792}
  (\bibinfo {year} {2008})}\BibitemShut {NoStop}%
\bibitem [{\citenamefont {Demir}\ and\ \citenamefont
  {Hanay}(2019)}]{demir2019}%
  \BibitemOpen
  \bibfield  {author} {\bibinfo {author} {\bibfnamefont {A.}~\bibnamefont
  {Demir}}\ and\ \bibinfo {author} {\bibfnamefont {M.~S.}\ \bibnamefont
  {Hanay}},\ }\href@noop {} {\bibfield  {journal} {\bibinfo  {journal} {IEEE
  Sensors Journal}\ } (\bibinfo {year} {2019})}\BibitemShut {NoStop}%
\end{thebibliography}%

	\clearpage
	
	\beginsupplement
	
	\section*{Supplementary}
	
	\subsection{Measurement setup details}
\begin{figure}[ht]
		\centering
		\includegraphics[width=0.5\textwidth]{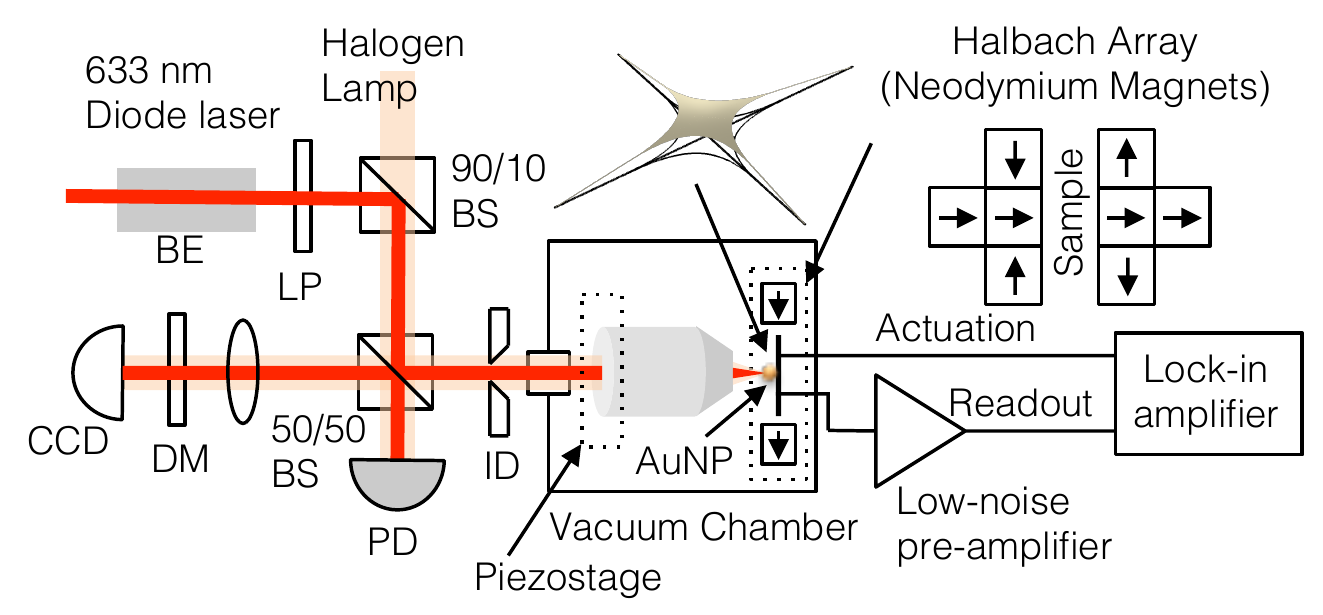} 
		\caption{Full schematics of the measurement setup. BE: beam expander. LP: linear polarizer. BS: beam splitter. PD: photodetector/powermeter. DM: dichroic mirror. ID: iris diaphram.}
		\label{fig:figS2} 
	\end{figure}

	The detailed measurement setup is shown in Figure~\ref{fig:figS2}. For better beam quality, a beam expander is used to match the spot size to the aperture of the objective. The beam is linearly polarized. The position of the beam is controlled by the piezostage with \SI{2}{\nano\meter} resolution and \SI{2}{\nano\meter} repeatability. All measurements are performed under a vacuum of \SI{1e-3}{\milli\Bar}. The enhanced Halbach array is used, as shown in Figure~\ref{fig:figS2}, with a gap of \SI{7}{\milli\meter} in between, resulting in averaged \SI{1}{\tesla} of static magnetic field in the center region of \SI{2}{\milli\meter} for efficient transduction.
	
\begin{figure}[ht]
		\centering
		\includegraphics[width=0.5\textwidth]{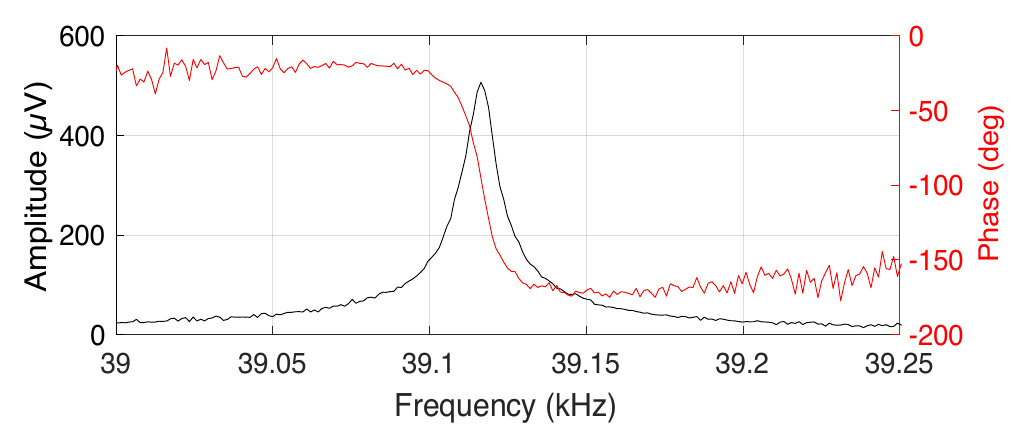} 
		\caption{Frequency sweep of the type C nanomechanical trampoline resonator with inductive readout and Lorentz force actuation obtained from a lock-in amplifier. The actuation voltage is \SI{10}{\micro\volt}. The \SI{-3}{\dB} bandwidth of the low-pass filter is \SI{5}{\hertz}.}
		\label{fig:figS1} 
	\end{figure}

	The inductive readout signal is first fed into a home-built low noise voltage pre-amplifier with a typical noise voltage of around \SI{1}{\nano\volt\per\sqrt{\hertz}} and a gain of around 200 before the lock-in amplifier and the phase-locked loop. Frequency sweep of a type C trampoline is shown in Figure~\ref{fig:figS1}. The clear phase signal at the resonance allows the frequency to be locked properly by the PLL.
	
	\subsection{Calculation of rise time}
\begin{figure}[ht]
		\centering
		\includegraphics[width=0.45\textwidth]{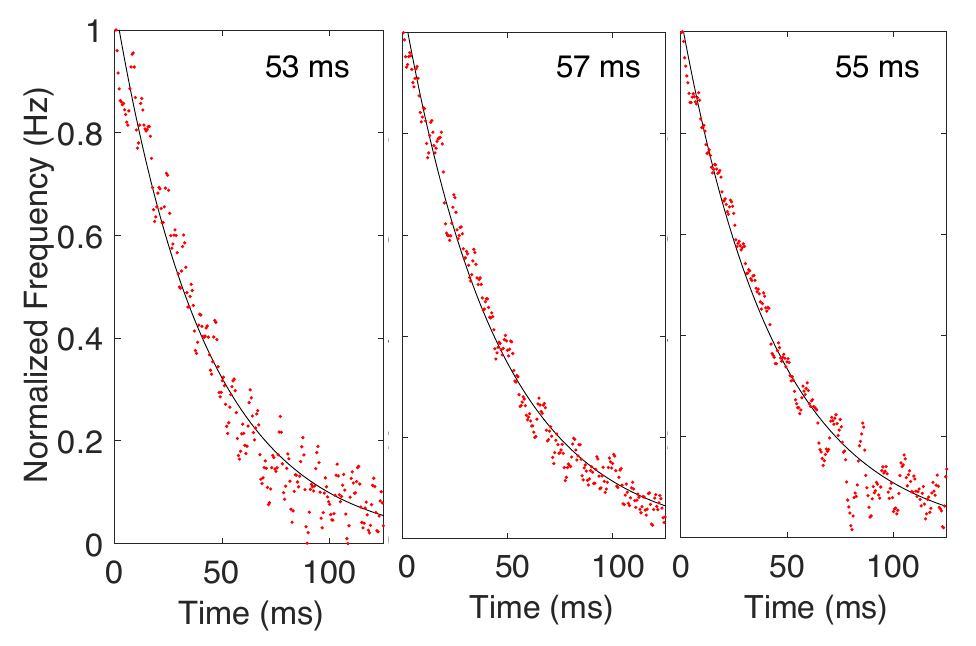} 
		\caption{The time constant of the NEMS-PSD subtracted by fitting the step frequency response with the first order low-pass model to obtain the rise time. These three step responses of frequency shift are sliced from Figure~\ref{fig:fig3}c.}
		\label{fig:figS3} 
	\end{figure}

	The rise time ($\tau_r$) of a step transition of frequency ($\Delta f$) can be calculated by the time difference between 10\% and 90\% transition. The rise time can also be related to the time constant ($\tau_{RC}$) of the first-order low-pass filter model of 
	\begin{equation}
	\Delta f = \Delta f_0 \big[1-exp(\frac{-t}{\tau_{RC}})\big]
	\end{equation}
	such that
	\begin{equation}
	\tau_r \approx 2.2 \tau_{RC}
	\end{equation}
	The extracted rise time from the first-order low-pass model of three different frequency steps in Figure~\ref{fig:fig3}c are shown in Figure~\ref{fig:figS3}.
	
	\subsection{Calculation of frequency noise for comparison with Allan deviation}
\begin{figure}[ht]
		\centering
		\includegraphics[width=0.45\textwidth]{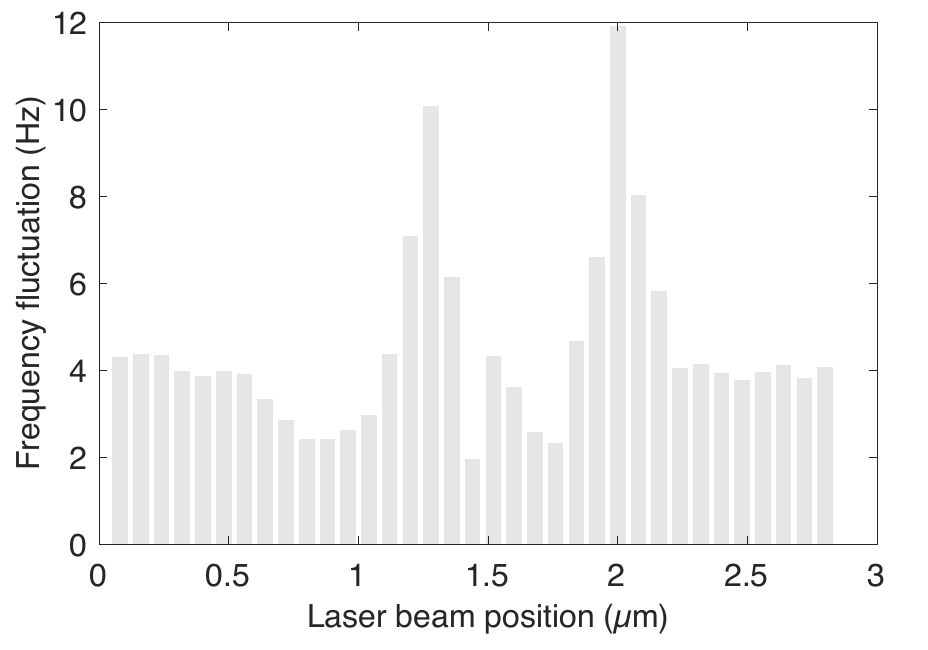} 
		\caption{Standard deviation of the resonance frequency of each step in Figure~\ref{fig:fig3}c.}
		\label{fig:figS4} 
	\end{figure}

	The standard deviation of each step in Figure~\ref{fig:fig3}c is first extracted as in Figure~\ref{fig:figS4}. Higher frequency fluctuation can be found in the larger step, due to slight oscillation of PLL resulting from the high instantaneous frequency change. The standard error ($\epsilon$) can be calculated from the standard deviation ($\sigma$) and the amount of samples of each step ($N$), such that
	
	\begin{equation}
	\epsilon = \frac{\sigma}{\sqrt{N}}
	\end{equation}
	
	In this measurement, the standard deviation of frequency of \SI{4}{\hertz} with around 250 samples each step results in a standard error of \SI{253}{\milli\hertz}, which is close to the Allan deviation measurement of \SI{250}{\milli\hertz} with same integration time of the step. If considering also the step with slight oscillation with an overall average standard deviation of \SI{6}{\hertz}, the standard error increases to \SI{380}{\milli\hertz}.
	
	\subsection{Calculation of theoretical responsivity}
	The calculation of theoretical responsivity is based on Equation~\ref{master}. The effective constant ($K_{eff}$) for the calculation is calculated based on the cross section ratio between 50 nm silicon nitride ($\sigma_{SiN}$) and 200 nm gold/chrome electrode($\sigma_{Au}$), such that
	\begin{equation}
	K_{eff} = \frac{K_{SiN} \times \sigma_{SiN} + K_{Au} \times \sigma_{Au}}{\sigma_{SiN}+\sigma_{Au}}
	\end{equation}
	K includes Young's modulus ($E$), thermal expansion coefficient ($\alpha$) and thermal conductivity ($\kappa$). The constants used for each material and the calculated effective constants are listed in the following table. The resulting displacement responsivity theoretical limit is calculated to be \SI{2080}{\hertz\per\micro\meter}.
	\begin{center}
		\begin{tabular}{|m{2em}|m{5em}|m{9em}|m{7em}|} 
			\hline
			& Young's modulus ($E$)  & Thermal expansion coefficient ($\alpha$) & Thermal conductivity ($\kappa$) \\
			\hline
			Au  & \SI{80}{\giga\pascal}  & \SI{2.2e-6}{\per\kelvin} & \SI{300}{\watt\per\meter\per\kelvin} \\ 
			\hline
			Cr  & \SI{280}{\giga\pascal} & \SI{5e-6}{\per\kelvin} & \SI{93}{\watt\per\meter\per\kelvin} \\ 
			\hline
			SiN & \SI{250}{\giga\pascal} & \SI{14e-6}{\per\kelvin} & \SI{3}{\watt\per\meter\per\kelvin} \\ 
			\hline
			eff. & \SI{179}{\giga\pascal} & \SI{7.2e-6}{\per\kelvin} & \SI{130}{\watt\per\meter\per\kelvin} \\
			\hline
		\end{tabular}
	\end{center}
	\subsection{Temperature field of trampoline}
\begin{figure}[ht]
		\centering
		\includegraphics[width=0.5\textwidth]{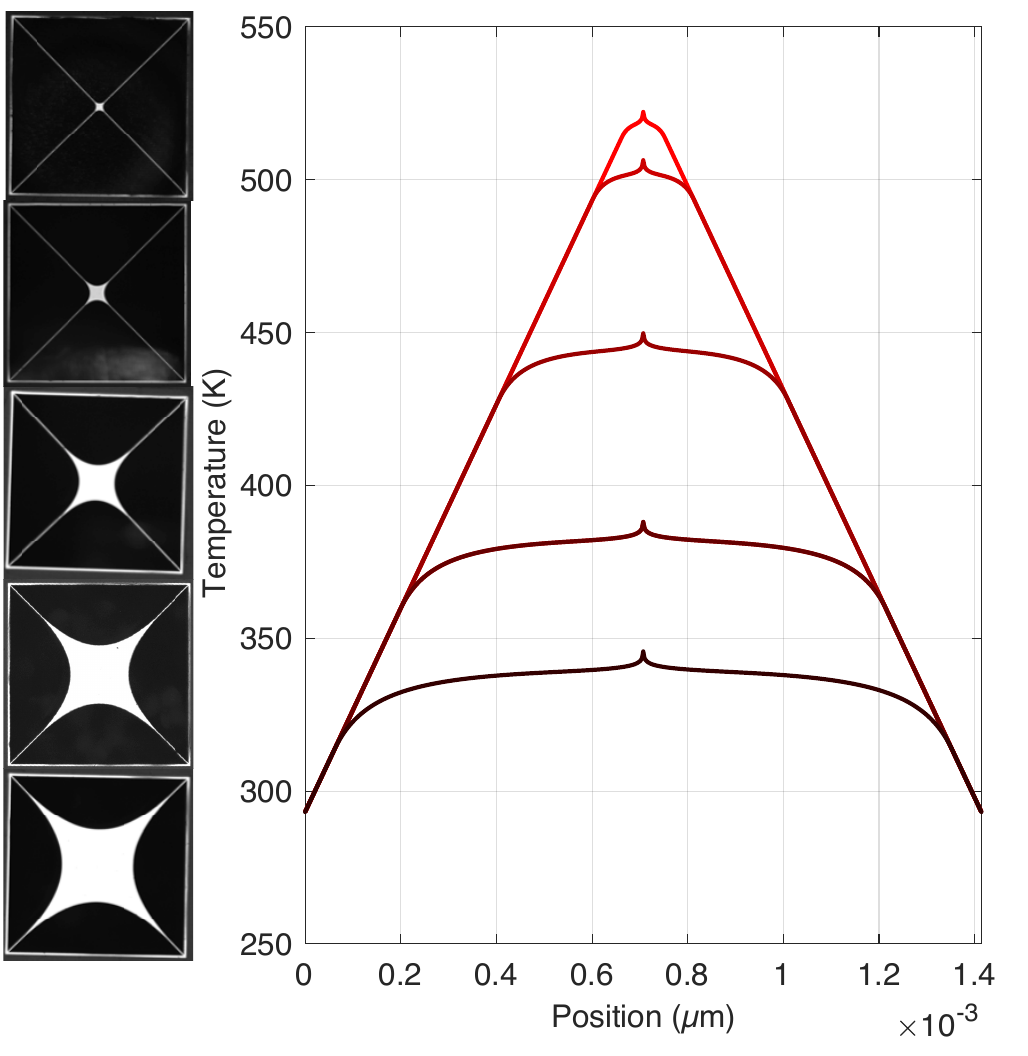} 
		\caption{Simulated temperature profile of trampoline resonators with different geometries. The point heating source is set to be \SI{1}{\micro\watt}}
		\label{fig:figS5} 
	\end{figure}

	The simulated temperature profile for different types of trampolines without electrodes are plotted in Figure~\ref{fig:figS5}. A point source of \SI{1}{\micro\watt} is put in the center of the structure. As the center area of the trampoline decreases, an increased overall temperature is created, and thus results in more stress relaxation and frequency shift. The temperature distribution shows a hybrid behavior of string and membrane. While the tether part has a linear temperature field like a string, the center part has exponentially decaying temperature field similar to a membrane.
	\subsection{Long-term stability}
\begin{figure}[ht]
		\centering
		\includegraphics[width=0.5\textwidth]{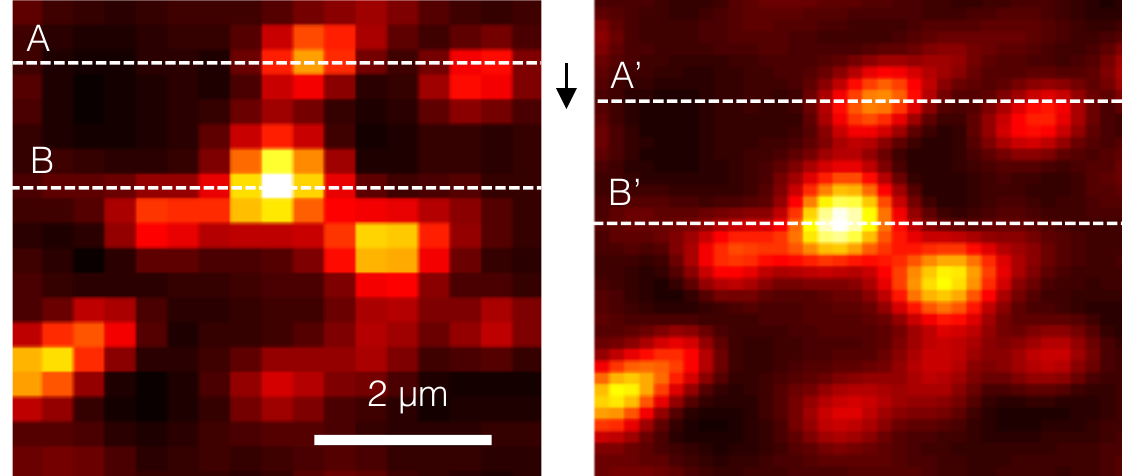} 
		\caption{Two scannings on the same area of a trampoline resonator distributed with gold nanoparticles with an interval of approximately 4 hours in between. After 4 hours, the reference line A and B drift slightly to line A' and B', as indicated in the black arrow, which corresponds to roughly 400 nm and 450 nm, respectively. The pixel size is different because of different scanning step, but this shouldn't influence the evaluation of the drift.}
		\label{fig:figS6} 
	\end{figure}

	Long-term stability is observed by repeating the scanning of same area with an interval of certain period of time, as shown in Figure~\ref{fig:figS6}.
	
\end{document}